\documentclass[journal=apchd5, manuscript=letter]{achemso}
\usepackage{achemso}
\usepackage[T1]{fontenc}
\usepackage{amsmath}
\usepackage{tabularx}
\usepackage{graphicx}
\usepackage{multirow}
\usepackage{ulem}
\usepackage{color}
\usepackage[draft]{fixme}
\usepackage{amssymb}
\usepackage{dcolumn}
\usepackage{bm}
\title{Stimulated emission depletion spectroscopy of color centers in hexagonal boron nitride}
\author{Ralph N. E. Malein}
\email{r.n.e.malein@exeter.ac.uk}
\affiliation{College of Engineering, Mathematics and Physical Sciences,University of Exeter, Exeter EX4 4QF, United Kingdom}
\author{Prince Khatri}
\affiliation{College of Engineering, Mathematics and Physical Sciences,University of Exeter, Exeter EX4 4QF, United Kingdom}
\author{Andrew J. Ramsay}
\affiliation{Hitachi Cambridge Laboratory, Hitachi Europe Ltd., Cambridge CB3 0HE, United Kingdom}
\author{Isaac J. Luxmoore}
\affiliation{College of Engineering, Mathematics and Physical Sciences,University of Exeter, Exeter EX4 4QF, United Kingdom}
\email{i.j.luxmoore@exeter.ac.uk}
\begin{document}

\begin{abstract}
We demonstrate the use of Stimulated Emission Depletion (STED) spectroscopy to map the electron-optical-phonon sideband of the ground state of the radiative transition of color centers in hexagonal boron nitride emitting at 2.0-2.2 eV, with in-plane linear polarization.
The measurements are compared to Photoluminescence of Excitation (PLE) spectra, that map the electron-optical-phonon sideband of the excited state. The main qualitative difference is a red-shift in the longitudinal optical phonon peak associated with $E_{1u}$ symmetry at the zone center. We argue that this is consistent with recent findings for a carbon-based line defect with admixture of energetically similar excited states.  
\end{abstract}

The progress of photonic quantum technologies hinges on the development of components such as quantum light sources and memories.  Due to their strong interaction with light and a wealth of fabrication and processing technology, atom-like solid state systems such as quantum dots or defects in wide-bandgap semiconductors hold much promise\cite{AtatureNatRevMat2018,AwschalomNatPhot2018}.  One area of particular promise are light-emitting defects in hexagonal boron nitride (hBN), which show bright emission of single photons with narrow linewidths, even at room temperature\cite{TranACSNano2016}, and with a low degree of emission into phonon-mediated modes\cite{TranNatNano2016, JungwirthNanoLetts2016}.  hBN's graphene-like two-dimensional structure allows straightforward incorporation into photonic devices\cite{GeimNature2013,liuNature2019,ProsciaArXiV2019}, and means that defects are always close to the device surface, which is desirable for the development of quantum sensors\cite{AnichiniChemSocRev2018,JimenezSciRep2020}.  Recently, optically-detected magnetic resonance (ODMR) has been observed in hBN, establishing the possibility to host spin qubits\cite{ChejanovskyArXiV2019,GottschollNatMat2020}. 

An open question in the field is the identity of the emitters. The issue is complicated by numerous candidate defects with similar zero-phonon line (ZPL) energies, motivating a search for additional spectroscopic signatures in ODMR \cite{ChejanovskyArXiV2019,GottschollNatMat2020}, or in the phonon sidebands \cite{linderalvArXiV2020,tawfikNanoscale2017,Wigger2DMat2019,GrossoACSPhot2020}.  For emitters at 2.0-2.2 eV, the strongest case to date has been made for a carbon-related defect \cite{MendelsonNatMat2020}.  

\begin{figure*}
\includegraphics[width=\textwidth]{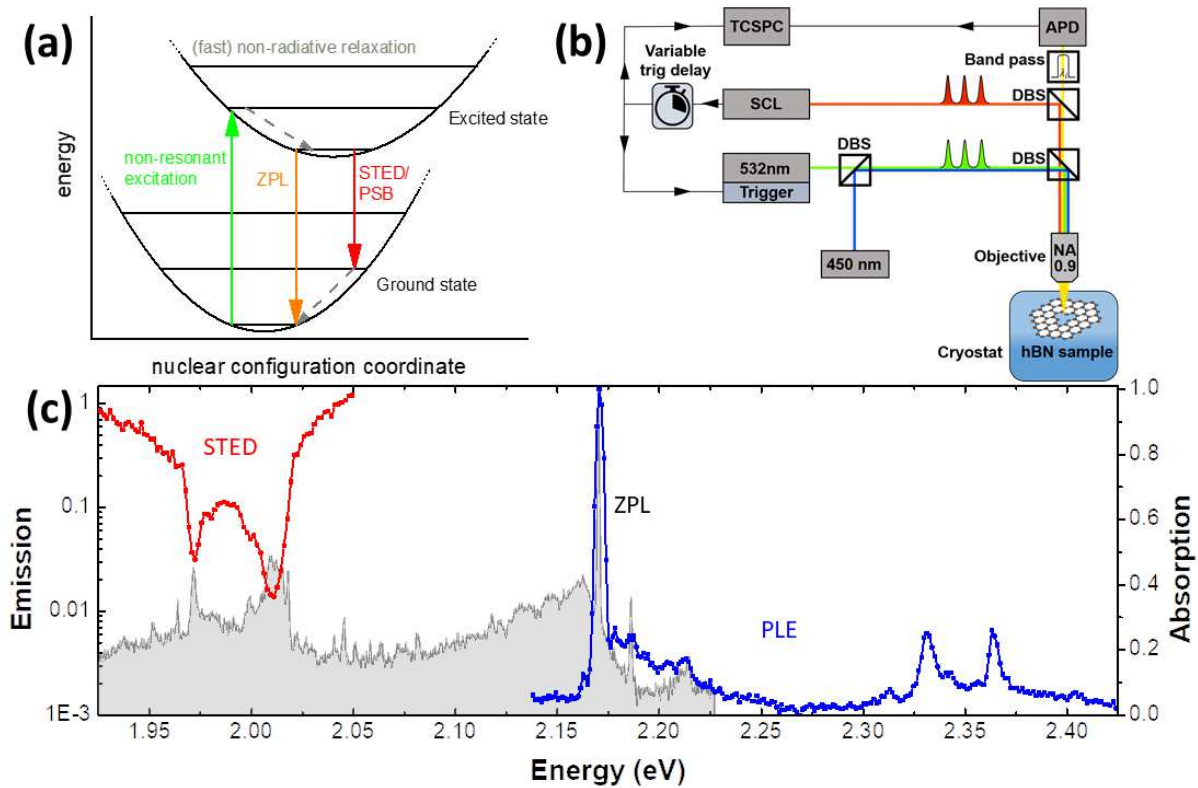}
\caption{\label{fig1}\textbf{(a)}: Franck-Condon energy diagram comparing PL, PLE, and STED techniques.  In PL, the non-resonant excitation (green) populates the excited state and emission from the ZPL (orange) and PSB (red) is collected.  In PLE, the excitation probes the excited vibronic states and the ZPL is collected.  In STED, the non-resonant excitation is again used, and the STED pulses (red) deplete the excited state through stimulated emission into the ground vibronic states, reducing the ZPL intensity. Grey dashed arrows show fast relaxation from higher vibronic states. \textbf{(b)}: Diagram of the experimental set-up. \textbf{(c)}: Representative PL (grey), PLE (blue) and STED (red) spectrum from defect-A with ZPL at $\sim$2.17eV, emission OPSB between 1.97 and 2.03eV and absorption OSPB between 2.32 and 2.38eV.  }
\end{figure*}

In this work we demonstrate stimulated emission depletion (STED) of color centers in hBN that emit around 2-2.2 eV. In STED an emitter is excited with two lasers with photon energies above and below the zero phonon line. The high energy laser pumps the emitter into the excited state, and the low energy laser stimulates phonon-assisted emission, thereby depleting the excited state and reducing the ZPL emission (Fig.~\ref{fig1}(a)). STED microscopy was developed for super-resolution spatial imaging\cite{HellOptLett1994,KlarOptLetts1999,BlomChemRev2017} and  has also been used to perform sub-resolution limit photolithography\cite{FischerOMExp2011,FischerAdvMat2012,FischerLPRev2013}, and to achieve lasing in NV centers in diamond\cite{JeskeNComms2016}. Here we apply STED as a spectroscopic probe, in combination with the complementary techniques of PL and PLE, to investigate the electron-phonon interaction of color centers in hBN. As illustrated in Fig.~\ref{fig1}(a), STED and PLE probe the vibronic manifold of the radiative ground and excited states, respectively. In all emitters studied, the 200-meV phonon peak corresponding to $\Gamma$-point of longitudinal optical phonon of $E_{1u}$ symmetry in bulk hBN \cite{SerranoPRL2007} is observed for the ground-state. However, for the excited state this peak is red-shifted. From this, and recent findings that defects emitting at this energy are likely carbon-boron substitutions with nitrogen vacancies $V_NC_B$\cite{MendelsonNatMat2020}, we infer that the excited state induces a distortion of the lattice in the plane that is not present in the ground state. This distortion results in a small local shift to the $LO(E_{1u})$ phonon mode energy.

Fig.~\ref{fig1}(a) illustrates the principle behind STED by comparison to photoluminescence excitation (PLE). In PLE, the system is driven by a laser at a higher energy than the ZPL, and the ZPL is collected. By tuning the excitation laser, the vibronic spectrum of the excited state can be measured. Conversely, in STED, the system is pumped into the excited state, in this case via non-resonant excitation as in photoluminescence (PL), and again the ZPL is collected. However, the system is then probed using a laser at a lower energy than the ZPL (red arrow).  As this probe laser is tuned, stimulated emission depletes the population of the excited state, reducing the intensity of the ZPL.  The PL signal against probe laser energy shows the vibronic spectrum of the ground state, where dips in the PL signal correspond to phonon resonances.  The STED spectrum replicates the PL spectrum, but without extraneous signals and peaks from adjacent defects or impurities. As such, STED is useful for PSB measurements of emitters with random placement, and in “dirty” systems such as two-dimensional materials where defects, impurities and surface effects are difficult to completely eliminate. 

The samples examined consist of few-layer flakes of hBN dropcast from solution onto a Si substrate coated with a 5nm layer of Al$_2$O$_3$.  The flakes are annealed at 850$^\circ$C for 15 minutes in a N$_2$ atmosphere to stimulate defect formation. Samples are then mounted in a closed-cycle cryostat and kept at 5 K. Fig.~\ref{fig1}(b) shows a schematic of the experimental set-up. The lasers used for excitation and depletion are collimated and co-aligned and then coupled to a long working distance objective lens, with numerical aperture of 0.8, which focuses the light to a diffraction-limited spot $< 1\mu$m in diameter.  Light emitted from the samples is collected into the same objective and coupled into a spectrometer and CCD for spectral measurements, or through a series of tuneable long- and short-pass filters for efficient wavelength selection, then to a single photon avalanche diode (SPAD) to perform photon counting and time-resolved fluorescence measurements via a time-tagging module. 

For non-resonant excitation in PL and STED measurements, a  green (532nm) pulsed laser  ($\sim$50 ps pulse width) is used.  In addition, a supercontinuum laser (SCL) fibre-coupled to an acousto-optic tuneable filter (AOTF) provides a tuneable pulsed excitation with a spectral range of 430 to 700nm and a bandwidth of ~1-2nm. The excitation is pulsed, with a repetition rate of 78MHz and a pulsewidth of a few ps. Depending on the spectral range selected this laser is used for both PLE and tuneable STED. To stabilize the PL from the defect, a weak 450nm blue CW-laser is also applied as in Ref.~\citenum{KhatriNL2020}. For STED measurements the green laser is triggered by a voltage pulse from the SCL, which has a tuneable delay, enabling control of the relative arrival time of excitation and depletion pulses.

In Fig.~\ref{fig1}(c), the PL spectrum of defect-A in hBN is plotted in grey, along with PLE (blue) and STED (red) spectra. To a first approximation, it is clear that the PLE spectrum is the mirror image of the PL spectrum around the ZPL energy, where PL (PLE) probes the emission (absorption) spectrum. However, in a system such as hBN, the PL spectrum can be contaminated by light from other nearby emitters. Applying STED by scanning a red-detuned laser shows resonances at the PSB but eliminates stray emission from other emitters.

\begin{figure*}
\includegraphics[width=\textwidth]{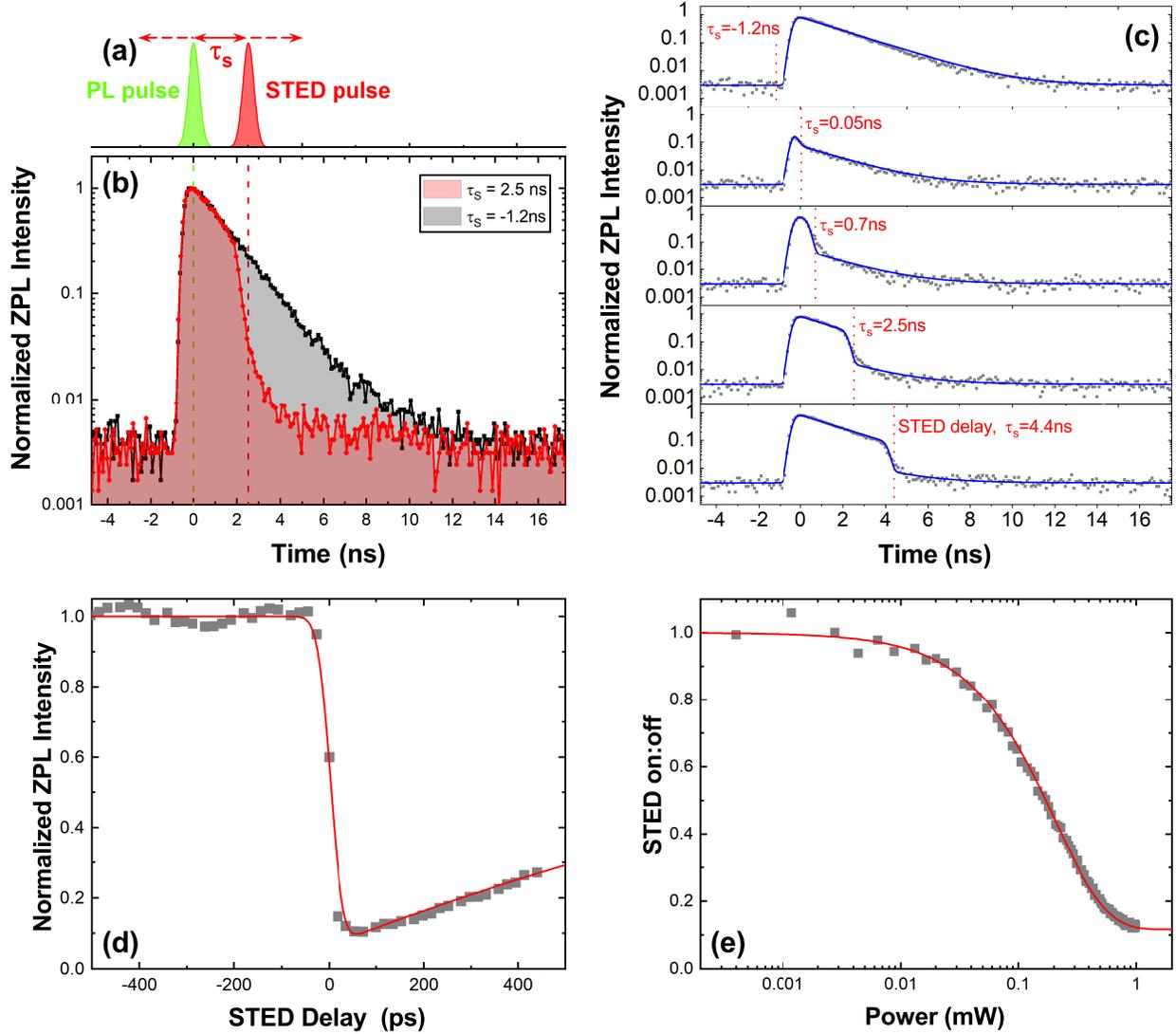}
\caption{\label{fig2}\textbf{(a)}: Schematic of arrival times of excitation PL pulse and STED pulse, with delay $\tau_s$ between them. \textbf{(b)}: Time-resolved ZPL-PL of defect-A with positive (red) and negative (black) $\tau_s$. The STED pulse switches off the emission \textbf{(c)}: Time resolved PL of defect-A at five different values of $\tau_s$, showing rapid depletion at arrival time of STED pulse. \textbf{(d)}: Plot of time-averaged ZPL-PL intensity with varying $\tau_s$.  Moving the STED pulse through the PL pulse results in significant quenching of the PL signal that recovers as $\tau_s$ is increased. \textbf{(e)}: Sweep of STED ratio with varying time-averaged power of the STED laser.}
\end{figure*}

To perform STED, time-resolved PL is recorded from the ZPL of defect-A shown in Fig.~\ref{fig1}(c). Following non-resonant excitation with a 532nm PL-pulse, the emission decays exponentially with radiative lifetime of 3.58ns, which is typical for hBN color centers. If the STED pulse, resonant with the PSB at $\sim$ 2.02eV, arrives before the excitation pulse, the PL is not affected (see black trace in Fig.~\ref{fig2}(b)).  However, if the STED pulse arrives after the excitation pulse, the PL is switched-off on the time-scale of the laser pulse (red trace in Fig.~\ref{fig2}(b)). The STED pulse stimulates phonon-assisted emission, depleting the excited state, and suppressing the PL from the ZPL.

The gating of the ZPL PL by the STED pulse is further illustrated in Fig.~\ref{fig2}(c), where time-resolved PL from defect-A are shown for five different values of $\tau_s$.  In Fig.~\ref{fig2}(d) the time-averaged PL  is plotted against delay time. With negative $\tau_s$ the PL intensity is constant and sharply falls as the pulses overlap, recovering slowly as the STED pulse is moved through the radiative decay tail. This provides a method of performing STED measurements without varying the SCL laser energy or switching lasers on and off, which can affect the power of the lasers and thus the reliability of measurements: ``STED on'' pulses are set to arrive 100ps after the excitation pulse, whereas ``STED off'' pulses arrive 100ps before the excitation pulse.  The STED on:off ratio is then recorded as the ratio between the STED on and off PL intensities.  A power sweep was performed, varying SCL power (Fig.~\ref{fig2}(e)).  Increasing SCL power decreases the STED ratio down to an apparent saturation at about 0.12. Experimental results were compared to simulations using a simple 3-level rate equation model and show good agreement (see Supporting Information for details on the model used).

\begin{figure*}
\includegraphics[width=300pt]{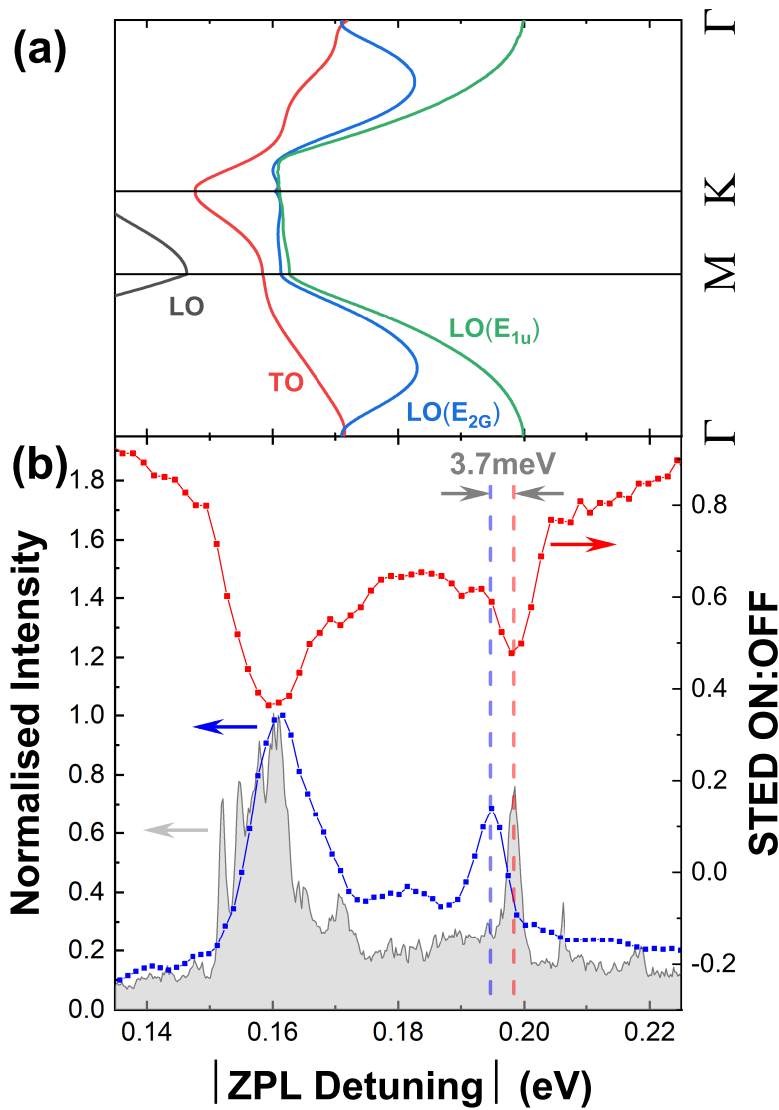}
\caption{\label{fig3}\textbf{(a)}: Calculated optical phonon dispersion for bulk hBN with $LO(E_{1u})$ mode highlighted. Taken from Serrano \textit{et al.} \cite{SerranoPRL2007}  \textbf{(b)}: Comparison of PL(grey), PLE (blue) and STED (red) OPSB spectra of defect-A, which has a ZPL energy of 2.170 eV. Red (blue) dashed line shows position of the $LO(E_{1u}, \Gamma)$ transition in STED (PLE).}
\end{figure*}

\begin{figure*}
\includegraphics[width=\textwidth]{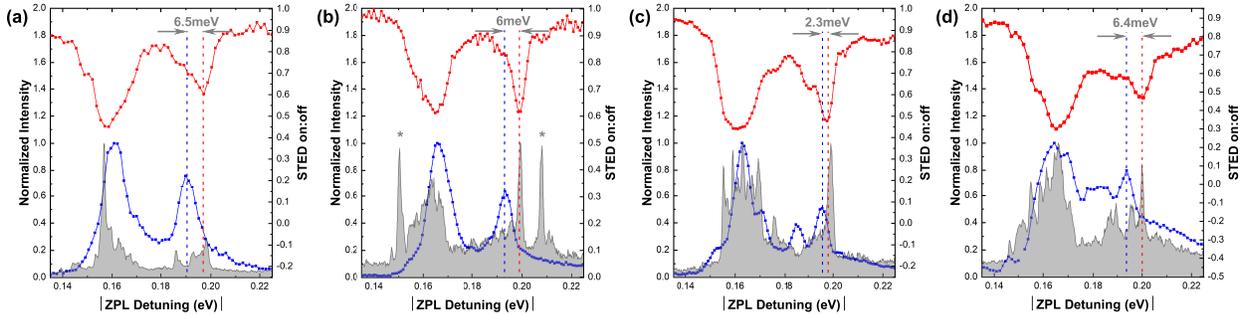}
\caption{\label{fig4}\textbf{(a)-(d)}: Comparison of PL, PLE and STED spectra from four similar defects with ZPL energies of (a) 2.166 eV (b) 2.175eV (c) 2.142 eV and (d) 2.171eV.  Each defect shows shift in 200meV peak between absorption and emission. In (b) the peaks marked with $\ast$ are from another nearby defect or defects.}
\end{figure*}

A comparison of STED, PL and PLE spectra for defect-A, along with four other defects with similar ZPL energy is made in Figs.~\ref{fig3} and~\ref{fig4}. For comparison, the magnitude of the detuning from the ZPL is used, because as noted above, PLE uses a laser detuned to higher energies, whereas STED uses a laser detuned to lower energies. As reported previously\cite{KhatriPRB2019, Vuong2DMat2016, VuongPRL2016}, the shape of the PSB corresponds closely to the phonon dispersion relation for 2-dimensional hBN (Fig.~\ref{fig3}(a))\cite{SerranoPRL2007}.  A detailed spectrum of the PL (emission) optical sideband (OPSB) along with corresponding PLE (absorption) data is shown in Fig.~\ref{fig3}(b).  The resolution of the PLE spectra is limited by the bandwidth of the SCL, and the PL shows sharper features. We note that the PL spectra has additional peaks not observed in PLE or STED. We attribute this to emission from other nearby emitters that are weakly excited. Hence, we only compare the PLE and STED spectra. The absorption (emission) PSB maps the coupling of the excited (ground) state of the radiative transition to the single phonon vibronic states. For defect-A (Fig.~\ref{fig3}), two main peaks are observed in both PLE and STED. The peak at $\sim$165meV corresponds directly to the maximum of the phonon density of states\cite{VuongPRL2016}, whereas the peak at 200 meV in STED and close to 200 meV in PLE arises from to Fr\"{o}hlich coupling to the longitudinal optical phonon mode with $E_{1u}$ symmetry at the $\Gamma$ point $LO(E_{1u}, \Gamma)$\cite{VuongPRL2016,KhatriPRB2019}. 

A shift in the energy of the $LO(E_{1u})$ phonon mode at the $\Gamma$-point implies that when the defect is in the excited state,  the lattice is distorted along a lattice coordinate with $E_{1u}$-like symmetry (in-plane dipole x,y), softening the spring-constant due to the anharmonicity of the bonds. We estimate this distortion to be about +2.6\% of the lattice constant (see Supporting Information for calculation). 

Recently, Mendelson {\it et al} \cite{MendelsonNatMat2020} have demonstrated that there is a carbon related defect emitting at $2.10\pm0.04~\mathrm{eV}$ with sharp ZPL, in-plane linear polarization, and nanosecond-scale radiative lifetimes, which matches our measurements.  Of the possible Carbon-related defects, the $(1)^4B_1 \rightarrow (1)^4A_2$ transition of the $V_NC_B^-$ line defect is the prime suspect. It has an in-plane linearly-polarized optical dipole perpendicular to the axis of the defect.  

In a previous work\cite{KhatriNL2020} on a similar defect to defect-A, we observe a second ZPL peak in absorption orthogonally polarised to and $\sim 0.5\pm 0.1$ eV above the emission ZPL, at 2.4-2.8 eV.  If the defect is $V_NC_B^-$ with $C_{2v}$ symmetry,  this should corresponds to the $(2)^4A_2 \leftrightarrow (1)^4A_2$ transition, which is calculated in supplement of ref.~\citenum{MendelsonNatMat2020} to have a vertical absorption energy between $2.8$ and $3.6$ eV.  Since this state does not appear in emission and was observed in PLE of the $\sim 2$eV ZPL, this indicates a fast non-radiative relaxation to the $(1)^4B_1$ state that flips the polarisation of the optical dipole, suggesting an intercrossing of the  $(2)^4A_2$ and $(1)^4B_1$ states. In this picture, the ground-state is energetically isolated with no in-plane dipole component, and should have a weak lattice distortion compared to the excited state where there are number of orbital states with similar energy available for admixing.  

In this work, we have demonstrated the application of  STED spectroscopy to the examination of vibronic states of defect emitters in hBN. We have shown that STED spectroscopy replicates the PL spectra, but with the advantage that STED completely eliminates stray signals from nearby emitters, making it immensely useful for systems with randomly-placed emitters in the solid state, such as defects in 2D materials or self-assembled quantum dots. We have shown that STED is analogous and complementary to PLE spectroscopy, where the major difference is that STED probes vibronic spectra of the ground state in a two-level system, whereas PLE probes that of the excited state.   We have then used STED and PLE to compare the vibronic spectra of the ground and excited states of the radiative transition. For color centers emitting near 2.2eV, the main qualitative difference between the ground and excited states is a red-shift in the LO-phonon mode with $E_{1u}$ symmetry. We compare our findings to recent work on the $V_NC_B^-$ defect and show that they are consistent with the $(1)^4B_1 \rightarrow (1)^4A_2$ transition. The shift to the phonon mode is ascribed to a lattice distortion due to admixing between nearby excited states in the defect.

Here, the STED resolution is limited to $\sim$1nm by the AOTF, which prevents a detailed examination of the OPSB around $165$ meV. However, with a narrower linewidth laser, PLE/STED would enable further investigation of the fine structure of the OPSB and thus shed more light on electron-phonon coupling in hBN defects.

\begin{acknowledgement}
This work was supported by the Engineering and Physical Sciences Research Council\\ (EP/S001557/1 and EP/026656/1).
\end{acknowledgement}

\begin{suppinfo}
\section{Model details}
To verify that STED is responsible for the experimental observations, we compare the results to a simple three-level rate equation model based on the level schematic in Fig.~\ref{figS1}.  $G$ and $E$ are the ground and excited state populations of the radiative transition respectively; $M$ is the intermediary vibronic level, lying above the ground state by the phonon mode energy. 

\begin{figure}
    \centering
    
    \includegraphics[width=0.6\textwidth]{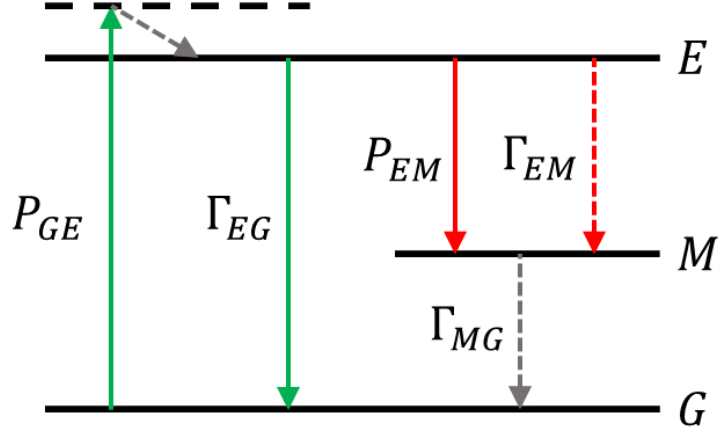}
    \caption{Level schematic of three-level system with rate parameters labeled.}
    \label{figS1}
\end{figure}

The excitation pulse is modelled as a single pump rate encompassing excitation from $G$ to a vibronic state above $E$ followed by rapid relaxation to $E$ ($P_{EM}$).  It is assumed that the non-radiative relaxation from $M$ to $G$ ($\Gamma_{MG}$) is fast compared to the other transitions. For time-resolved simulations, the excitation and STED pulses are modelled as Gaussian, and the simulated traces are then convolved with the IRF of the SPAD (Gaussian with 800 ps FWHM) to better fit to the measurement resolution. The ZPL intensity is then proportional to the population in $E$ multiplied by the radiative decay rate $\Gamma_{EG}$.  A phonon-mediated spontaneous emission rate $\Gamma_{EM}$ is also included to model emission into the PSB. The rate equations are:

\begin{align*}
    \dot{G} &= \Gamma_{EG} E + \Gamma_{MG} M - P_{GE} G \\
    \dot{E} &= P_{GE} G - \left(\Gamma_{EG} + P_{EM} + \Gamma_{EM}\right) E\\
    \dot{M} &= -(\dot{E} + \dot{G})
\end{align*}

All simulations show good agreement with experimental data with consistent model parameters (see Table~\ref{table1}): fitting to the time-resolved PL traces in Fig.~2(c) gives values of the radiative decays and allows accurate determination of the true pulse widths of the 532nm laser and SCL pulses; fitting the power sweep in Fig.~2(e) shows that the saturation is due to the short delay between excitation and STED pulses which allows a small degree of PL to be emitted before depletion of $E$.

\begin{table}[ht]
    \centering
    \begin{tabular}{c|c}
        model parameter & value \\
        \hline
        $\Gamma_{EG}$, $\Gamma_{EM}$ & 279.3 MHz \\
        $\Gamma_{MG}$ & 1 THz \\
        $P_{GE}$ & 8 GHz \\
        $P_{EM}$ in delay sweep (Fig.~2(d)) & 1.29 THz \\
        Excitation laser pulse width & 50 ps \\
        STED pulse width & 2 ps
    \end{tabular}
    \caption{Table of simulation parameters for 3-level model shown in Fig.~\ref{figS1}.}
    \label{table1}
\end{table}

\section{Estimation of lattice distortion}

The distortion in the lattice is estimated by assuming a softening of the spring constant $K$ of the phonon mode due to the distortion.  For simplicity's sake, we consider a 1D diatomic chain whose dispersion is given by
\begin{equation*}
    \omega^2 = K \frac{m_B + m_N}{m_Bm_N} \pm \sqrt{K^2 \left(\frac{(m_B + m_N)^2}{m_B^2m_N^2} - \frac{4}{m_Bm_N}\sin^2\left(\frac{k a}{2}\right)\right)}
\end{equation*}
where $k$ is the phonon momentum, $a$ is the lattice parameter, and $m_{B,N}$ are the masses of the two atoms.  As the interaction take place at the $\Gamma$ point $k=0$, the dispersion simplifies to 
\begin{align*}
    \omega^2 &= K \frac{m_B + m_N}{m_Bm_N} \pm \sqrt{K^2 \frac{(m_B + m_N)^2}{m_B^2m_N^2}} \\
     &= 0 \quad \mathrm{ or }\quad 2K \frac{m_B + m_N}{m_Bm_N} 
\end{align*}
Thus in the optical branch, $\omega^2 \propto K$.  While hBN is not a diatomic chain, considering units of $K$, this proportionality holds for a 2D hexagonal lattice.

Considering Hooke's law 
\begin{equation*}
    K \propto \left. \frac{\partial^2 V}{\partial q^2}\right|_{q = q_0} 
\end{equation*}
where $V$ is the bond potential, $q$ is the normal mode lattice coordinate and $q_0$ is the equilibrium lattice coordinate. Assuming a form of the potential similar to the Lennard-Jones potential, where it is minimized at the equilibrium position, and expanding around the equilibrium, we find that 
\begin{align*}
    \left. \frac{\partial^2 V}{\partial q^2}\right|_{q = q_0} &\propto \frac{1}{q_0^2} \\
    &\mathrm{so} \\
    \omega &\propto \frac{1}{q_0}
\end{align*}
Thus, considering excited (ground) phonon mode energies $\hbar\omega_E = 195 \pm 2$meV ($\hbar\omega_G = 200$meV) and equilibrium positions $q_E$ ($q_G$) we can calculate the excited state lattice equilibrium position as a percentage of the ground state equilibrium:

\begin{align*}
    \frac{q_E}{q_G} &= \frac{\hbar\omega_G}{\hbar\omega_E} \\
    &= \frac{200}{195\pm2}\\
    &= 102.6\pm 1\%
\end{align*}

Leading to an overall distortion of $+2.6 \pm 1\%$.

\end{suppinfo}

\bibliography{hBN_STED_spectroscopy}

\end{document}